\documentclass[12pt]{article}
\usepackage{epsfig}
\usepackage{amsfonts}

\newcommand{\Li}{\mbox{Li}}
\newcommand{\ep}{\epsilon}
\newcommand{\be}{\beta}
\newcommand{\al}{\alpha}

\newcommand\VV{\setbox0=\hbox{V}\hbox{\rm V\raise\ht0
  \hbox to0pt{\hss\vbox to0pt{\hbox{v}\vss}}}}
\def\slashchar#1{\setbox0=\hbox{$#1$}           
   \dimen0=\wd0                                 
   \setbox1=\hbox{/} \dimen1=\wd1               
   \ifdim\dimen0>\dimen1                        
      \rlap{\hbox to \dimen0{\hfil/\hfil}}      
      #1                                        
   \else                                        
      \rlap{\hbox to \dimen1{\hfil$#1$\hfil}}   
      /                                         
   \fi}                                         %

\begin{document}

\begin{center}
{\Large \bf $\rho$-meson form factors and QCD sum rules.}
\\ \vspace*{5mm} V.V.Braguta$^{a}$ and A.I.Onishchenko$^{b}$
\end{center}

\begin{center}
a) Institute for High Energy Physics, Protvino, Russia \\
\vspace*{0.5cm} 
b) Department of Physics and Astronomy \\ Wayne State University,
Detroit, MI 48201, USA \\
\end{center}

\vspace*{0.5cm}

\begin{abstract}
We present predictions for $\rho$-meson form factors obtained from the analysis
of QCD sum rules in next-to-leading order of perturbation theory. The radiative
corrections turn out to be sizeable and should be taken into account
in rigorous theoretical analysis. 
\end{abstract}

\section{Introduction}

The method of QCD sum rules \cite{QCDSR} is designed to estimate low-energy characteristics of hadrons,
such as masses, decay constants and form factors. Within this framework we analyze the correlation
function of currents in deep euclidean region with the help of operator product expansion, which
allows us to take into account both perturbative and nonperturbative contributions. The presence 
of latter could be traced to the non-vanishing values of vacuum QCD condensates. Physical quantities,
we are interested in, are determined by matching this correlator to its phenomenological representation.     

In this work we performed an analysis of three-point sum  rules for $\rho$-meson form factors at
intermediate momentum transfer. Basically, it is an extension of already available LO analysis
\cite{Ioffe:qb} to include radiative corrections. To compute radiative corrections we used the technic
already developed and tested in the analysis of pion electromagnetic form factor within NLO QCD sum rule setup
both with pseudoscalar and axial-vector pion interpolating currents \cite{Braguta:2003hd,Braguta:2004ck}.

The paper is organized as follows. In section 2 we describe our framework and give explicit
expressions for next-to-leading order corrections to double spectral density. Section 3
contains our numerical analysis and expressions for the contributions
of gluon and quark condensates. Finally, in section 4 we draw our conclusions.

\section{The method}

To determine $\rho$-meson electromagnetic form factors we will use the method of three-point
QCD sum rules. Within this framework $\rho$-meson is described as a result of an action of vector
interpolating current on vacuum state. We define the vacuum to $\rho$-meson transition matrix element
of vector current as
\begin{eqnarray}
\langle 0|j_{\mu}|\rho^+,\ep\rangle = \frac{m_{\rho}^2}{g_{\rho}}\ep_{\mu},
\end{eqnarray}
where $m_{\rho}$ is the $\rho$-meson mass, $g_{\rho}$ is the $\rho -\gamma$ coupling constant 
($g_{\rho}^2/4\pi = 1.27$) and $\ep_{\mu}$ stands for $\rho$-meson polarization vector. Next, 
assuming parity and time-reversal invariance, the general expression for $\rho$-meson  electromagnetic
vertex could be written in terms of three form factors:
\begin{eqnarray}
\langle\rho^+ (p',\ep')| j_{\mu}^{\mathbf{el}}|\rho^+ (p,\ep)\rangle &=& -\ep_{\be}^*\ep_{\al}
\Biggl\{
\left[(p'+p)_{\mu}g_{\al\be} - p'_{\al}g_{\be\mu} - p_{\be}g_{\al\mu} \right]F_1 (Q^2)
\Biggr. \nonumber \\
&& \left. 
+ [g_{\mu\al}q_{\be} - g_{\mu\be}q_{\al}]F_2 (Q^2) + \frac{1}{m_{\rho}^2}p'_{\al}p_{\be}(p+p')_{\mu}F_3 (Q^2)
\right\},
\end{eqnarray}
where $j_{\mu}^{\mathbf{el}} = e_u\bar u\gamma_{\mu}u + e_d\bar d\gamma_{\mu}d$, the momenta of initial and
final state $\rho$-mesons were denoted by $p$, $p'$ and $Q^2 = -q^2 (q=p-p')$ is the square of momentum transfer.
$F_1$, $F_2$ and $F_3$ are electric, magnetic and quadrupole form factors. At zero momentum transfer these form factors
are expressed through the usual static quantities of $\rho$-meson charge, magnetic moment ($\mu$) and quadrupole 
moment ($D$):
\begin{eqnarray}
F_{1} (0) &=& 1, \\
F_{2} (0) &=& \mu - 1, \\
F_{3} (0) &=& \frac{1}{2}(\mu - 1) - \frac{m_{\rho}^2}{4}D.
\end{eqnarray}  
Within approach of QCD sum rules the theoretical estimates of $\rho$-meson form factors follow from the
analysis of the following three-point correlation function:
\begin{eqnarray}
\Pi_{\mu\alpha\beta} = i^2 \int d^4 x d^4 y e^{i(p'\cdot x- p\cdot y)} 
\langle 0| T\{j_{\be}^{+}(x), j_{\mu}^{\mathbf{el}}(0), j_{\al}(y) \}|0\rangle, \label{correlator}
\end{eqnarray}   

\begin{figure}[ht]
\begin{center}
\includegraphics[scale=0.5]{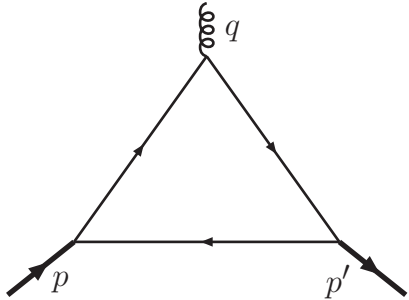} 
\caption{LO diagram}
\label{LOdiagram}
\end{center}
\end{figure}

The scalar amplitudes $\Pi_i$ in front of different Lorentz structures are the functions of kinematical invariants, 
i.e. $\Pi_i = \Pi_i (p^2,p'^2,q^2)$. In the  region of large euclidean momenta $p^2,p'^2,q^2 < 0$ this correlation 
function could be studied with the use
of ordinary perturbative QCD. The calculation of QCD expression for three-point correlator is done through the use 
of operator product expansion (OPE) for the T-ordered product of currents. As a result of OPE one obtains besides 
leading perturbative contribution also power corrections, given by vacuum QCD condensates. We will return to
the discussion of power correction to QCD sum rules later after the definition of Borel transform for our correlation
function. Now let us discuss the perturbative contribution. The calculation of perturbative contribution could be 
conveniently performed with the use of double dispersion representation in variables $s_1 = p^2$ and $s_2 = p'^2$
at $q^2 < 0$:
\begin{eqnarray}
\Pi_{\mu\alpha\beta}^{\mathbf{pert}}(p^2, p'^2, q^2) = \frac{1}{(2\pi)^2}
\int\frac{\rho_{\mu\alpha\beta}^{\mathbf{pert}} (s_1,s_2,Q^2)}{(s_1-p^2)(s_2-p'^2)} ds_1ds_2 + \mbox{subtractions}
\label{doubledisp} 
\end{eqnarray} 
The integration region in (\ref{doubledisp}) is determined by condition
\footnote{In our case this inequality is satisfied identically.}
\begin{eqnarray}
-1 \leq \frac{s_2 - s_1 -q^2}{\lambda^{1/2} (s_1, s_2, q^2)} \leq 1
\end{eqnarray}
and
\begin{eqnarray}
\lambda (x_1, x_2, x_3) = (x_1 + x_2 -x_3)^2 - 4 x_1 x_2.
\end{eqnarray}
The double spectral density $\rho_{\mu\alpha\beta}^{\mathbf{pert}}(s_1,s_2,Q^2)$ is searched in the form of expansion
in strong coupling constant:
\begin{eqnarray}
\rho_{\mu\alpha\beta}^{\mathbf{pert}}(s_1,s_2,Q^2) = 
\rho_{\mu\alpha\beta}^{(0)}(s_1,s_2,Q^2) + \left(\frac{\alpha_s}{4\pi}\right)\rho_{\mu\alpha\beta}^{(1)}(s_1,s_2,Q^2)
+ \ldots
\end{eqnarray}

\begin{figure}[ht]
\begin{center}
\includegraphics[scale=0.3]{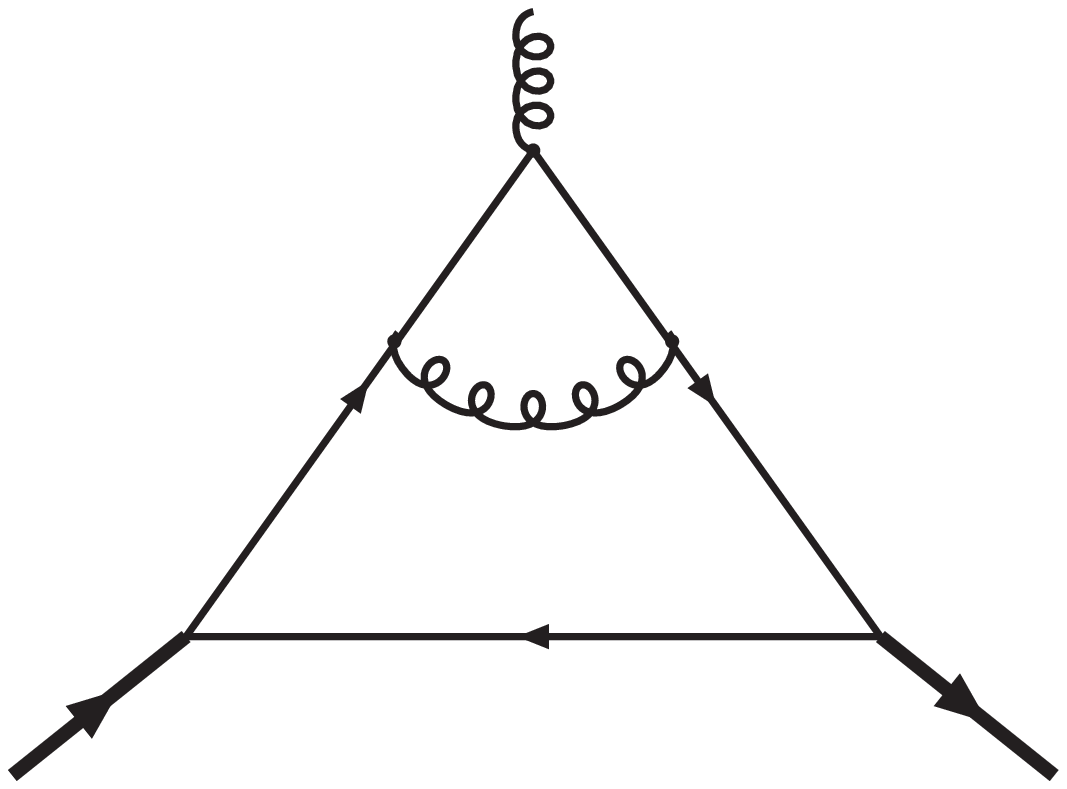} \makebox[2.cm]{}
\includegraphics[scale=0.3]{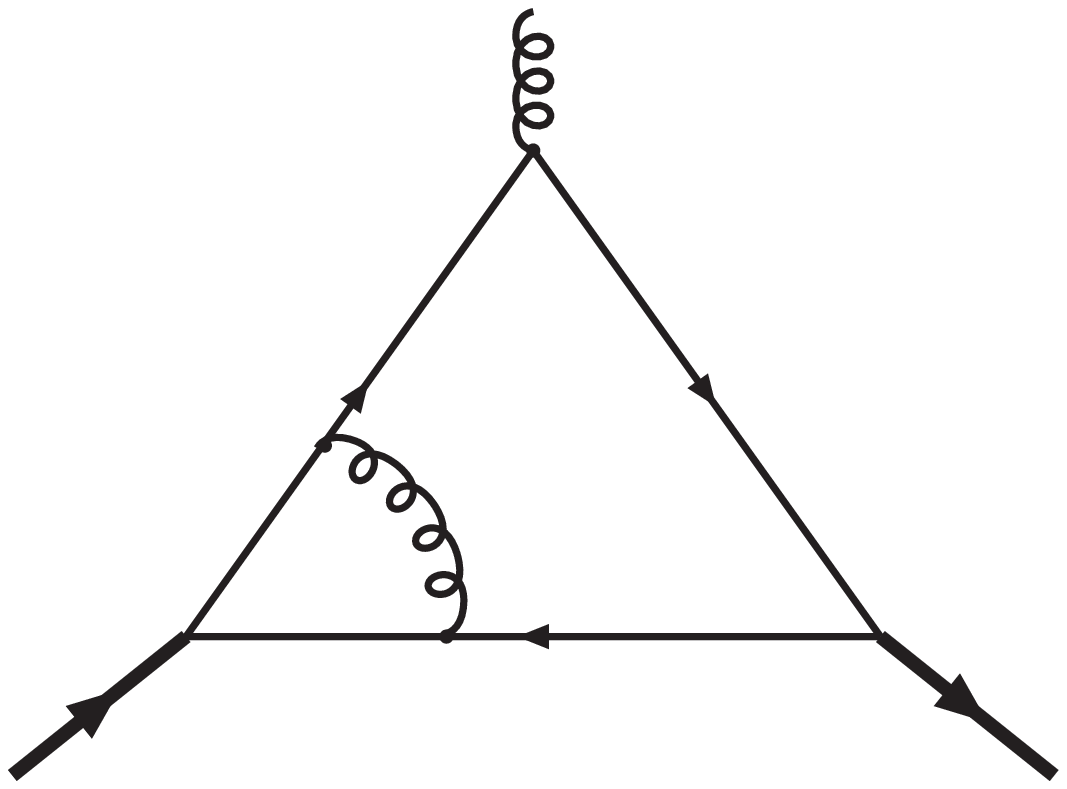} \makebox[2.cm]{} 
\includegraphics[scale=0.3]{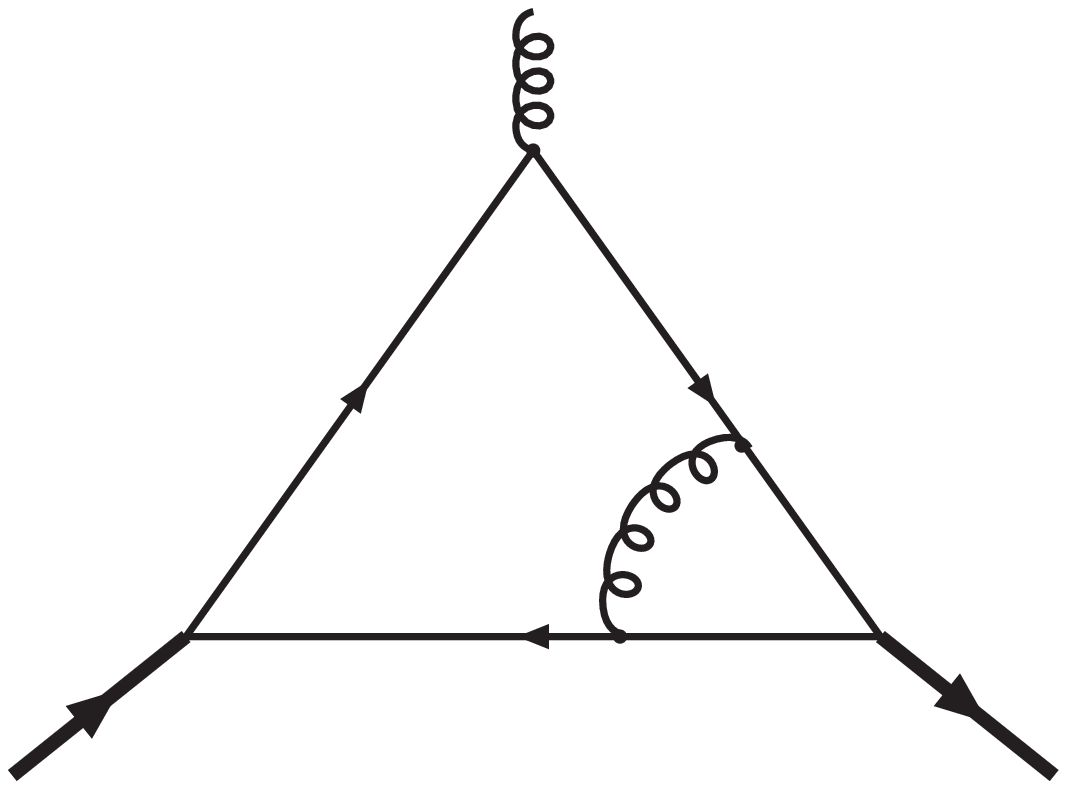} \\ \vspace*{1cm}
\includegraphics[scale=0.3]{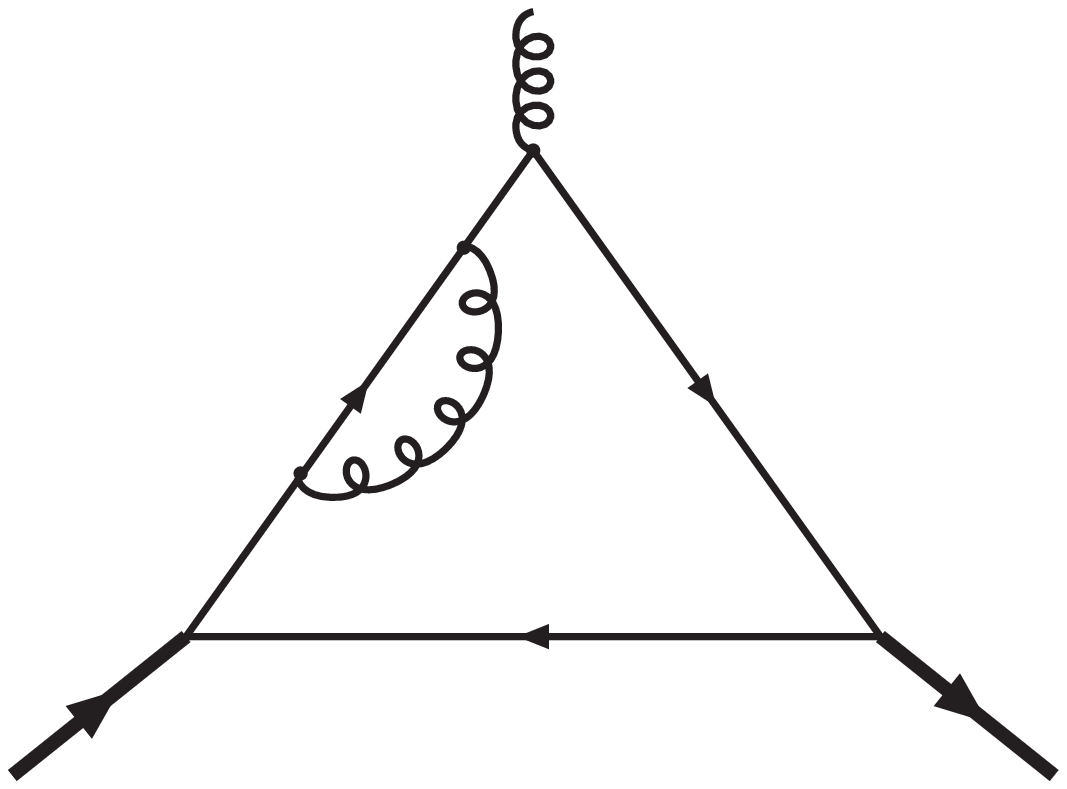} \makebox[2.cm]{}
\includegraphics[scale=0.3]{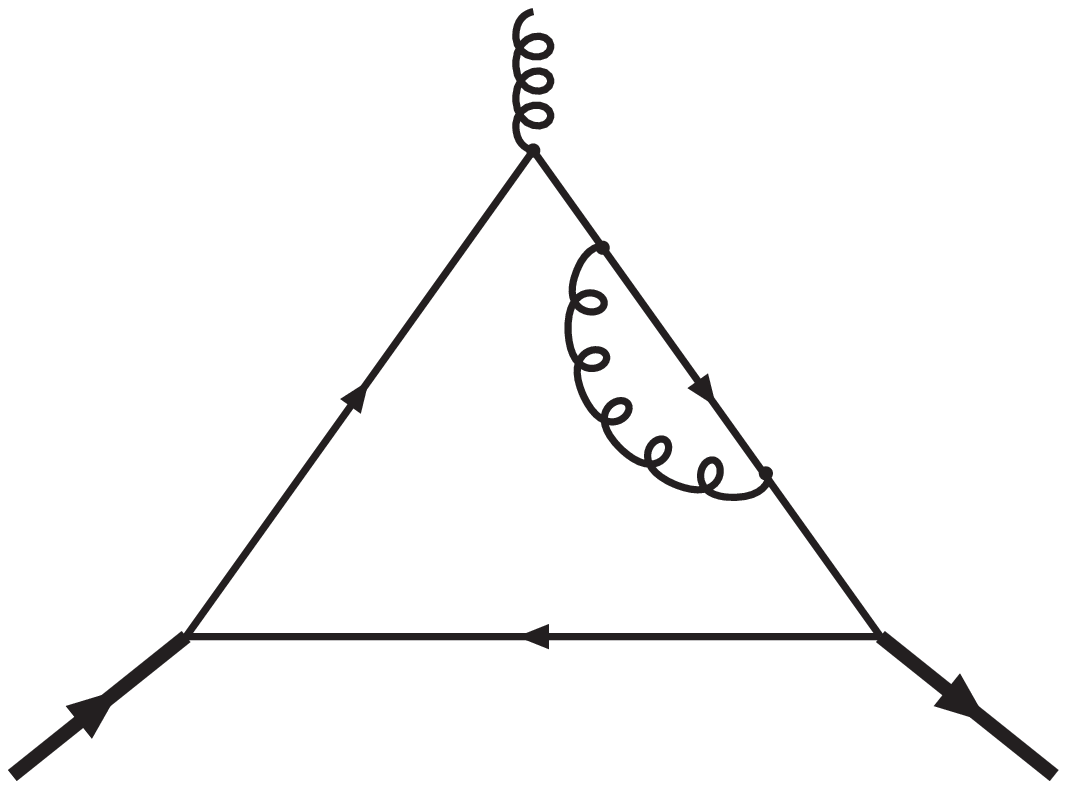} \makebox[2.cm]{} 
\includegraphics[scale=0.3]{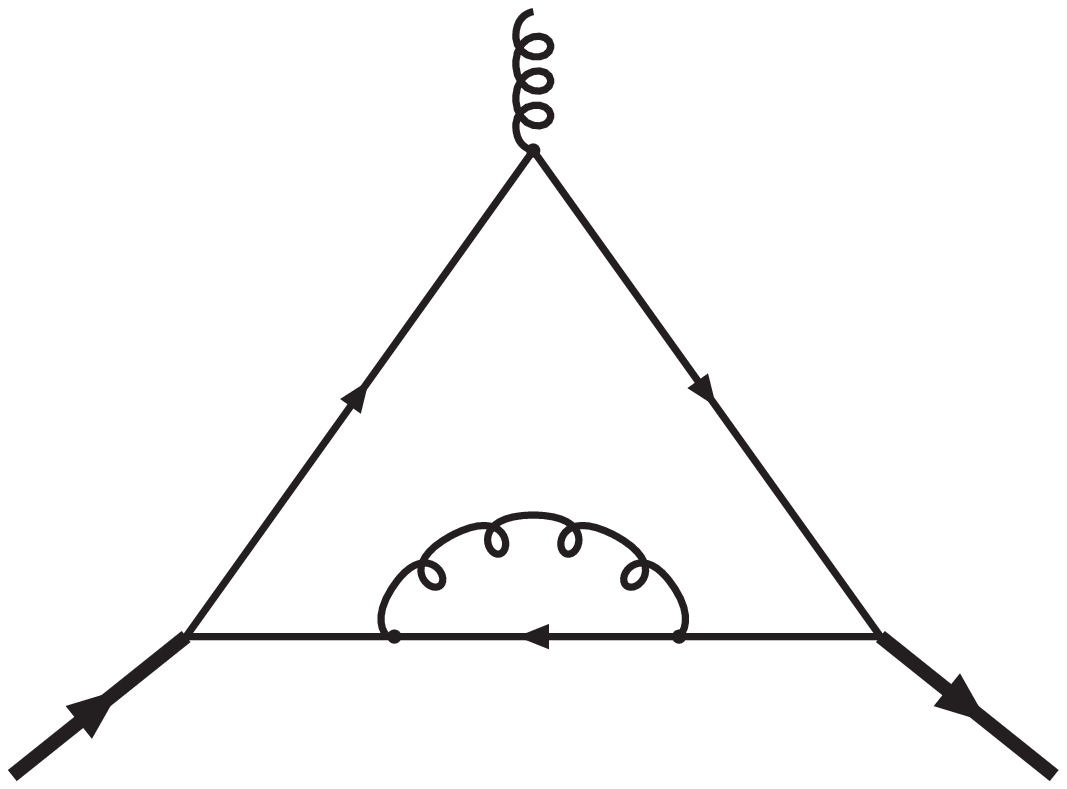}
\vspace*{0.5cm}
\caption{NLO diagrams}
\label{NLOdiagrams}
\end{center}
\end{figure}

At leading order in coupling constant we have only one diagram depicted in Fig. 1, contributing to
three-point correlation function. At next to leading order we have 6 diagrams shown in Fig. 2.
The calculation of corresponding double spectral density was performed with the standard use
of Cutkosky rules. In the kinematic region $q^2 < 0$, we are interested in, there is no 
problem in applying Cutkosky rules for determination of $\rho_{\mu\alpha\beta} (s_1, s_2, Q^2)$ and integration
limits in $s_1$ and $s_2$. The non-Landau type singularities, not accounted for by Cutkosky
prescription, do not show up here. The calculation could be considerably simplified with
the use of Lorentz decomposition of double spectral density based on a fact, that our spectral density
is subject to three transversality conditions: 
$\rho_{\mu\al\be}q_{\mu} = \rho_{\mu\al\be}p_{\al} = \rho_{\mu\al\be}p'_{\be} = 0$
\begin{eqnarray}
\rho^{\mu\alpha\beta} &=&  A_1 [(Q^2+x)p_1^{\alpha}-(x+y)p_2^{\alpha}]
[(y-x)p_1^{\beta}+(Q^2+x)p_2^{\beta}][(Q^2+y)p_1^{\mu}+(Q^2-y)p_2^{\mu}] \nonumber \\
&& - \frac{1}{2}A_2 [(Q^2+y)p_1^{\mu}+(Q^2-y)p_2^{\mu}][(Q^2+x)g^{\alpha\beta}-2p_1^{\beta}p_2^{\alpha}] \nonumber \\
&& - \frac{1}{2}A_3 [(Q^2+x)p_1^{\alpha}-(x+y)p_2^{\alpha}][2(p_2^{\beta}-p_1^{\beta})p_2^{\mu}+(Q^2+y)g^{\mu\beta}]
\nonumber \\
&& - \frac{1}{2}A_4 [(x-y)p_1^{\beta}-(Q^2+x)p_2^{\beta}][2(p_2^{\alpha}-p_1^{\alpha})p_1^{\mu}+(y-Q^2)g^{\mu\alpha}],
\end{eqnarray}
where $x=s_1+s_2$, $y=s_1-s_2$ and $p_1 = p, p_2 =p'$. The four independent structures 
$A_i$ (we suppressed the dependence on kinematical invariants) are given by a solution of system 
of linear equations: 
\begin{eqnarray}                
I_1 &=& \rho_{\mu \alpha \beta} p_1^{\mu} p_2^{\alpha} p_1^{\beta} = 
\frac {k^2} 8 \biggl ( k A_1 - A_2 - A_3 - A_4 \biggr )
\\
I_2 &=& \rho_{\mu \alpha \beta} p_1^{\mu} g^{\alpha \beta} = 
\frac{k}{4} (x+Q^2) \biggl ( k A_1 - 3 A_2 - A_3 - A_4 \biggr )
\\ 
I_3 &=& \rho_{\mu \alpha \beta} p_2^{\alpha} g^{\mu \beta} = 
\frac{k}{4} (y+Q^2) \biggl ( k A_1 - A_2 - 3 A_3 - A_4 \biggr )
\\ 
I_4 &=& \rho_{\mu \alpha \beta} p_1^{\beta} g^{\mu \alpha} = 
- \frac {k}{4} (y-Q^2) \biggl ( k A_1 - A_2 - A_3 - 3 A_4 \biggr ),
\label{system}
\end{eqnarray}         
where $k = \lambda (s_1,s_2,-Q^2)$.
Solving this system it is easy to find explicit expressions for $A_i$ in terms of $I_i$ 
(functional dependence on kinematical invariants is assumed):
\begin{eqnarray}
A_1 &=& \frac{20}{k^3}I_1 - \frac{2}{(Q^2+x)k^2}I_2 - \frac{2}{(Q^2+y)k^2}I_3 - \frac{2}{(Q^2-y)k^2}I_4, \\
A_2 &=& \frac{4}{k^2}I_1 - \frac{2}{(Q^2+x)k}I_2, \\
A_3 &=& \frac{4}{k^2}I_1 - \frac{2}{(Q^2+y)k}I_3, \\
A_4 &=& \frac{4}{k^2}I_1 - \frac{2}{(Q^2-y)k}I_4. 
\end{eqnarray}
At Born level and expressions for $I_i$ are easy to find and they are given by ($s_3 = Q^2$) \cite{Ioffe:qb}:
\begin{eqnarray}
I_1^{(0)} &=& -\frac{3s_1s_2s_3}{2k^{1/2}}, \\
I_2^{(0)} &=& -\frac{3s_1s_2}{k^{1/2}}, \\
I_3^{(0)} &=& -\frac{3s_2s_3}{k^{1/2}}, \\
I_4^{(0)} &=& -\frac{3s_1s_3}{k^{1/2}}, 
\end{eqnarray}
 The calculation of NLO radiative corrections to double
spectral density is in principle straightforward. One just needs to consider all possible
double cuts of diagrams, shown in Fig. 2. However, the presence of collinear and soft infrared
divergences calls for appropriate regularization of arising divergences 
at intermediate steps of calculation and makes the whole analytical calculation quite
involved. We will present the details of NLO calculation in one of our future publications. Here
we give only final results:
\begin{eqnarray}
k^{1/2}I_1^{(1)} &=& -s_1^3 + s_2s_1^2 + s_2^2s_1-s_2^3 + (s_1+s_2)s_3^2 - s_1^2s_3 - s_2^2s_3 + s_1s_2s_3\Bigl[ \Bigr.\nonumber \\ 
&& - 16\log^2(v_1) - 16\log (v_3)\log (v_1) -16\log (v_4)\log (v_1)  \nonumber \\
&& + 2\log (v_1) - 4\log^2 (v_3) - 4\log^2 (v_4) - 2\log (v_2) - 2\log (v_3)
- 8\log (v_3)\log (v_4) \nonumber \\ && \left.
-8\Li_2\left(\frac{x_2}{x_1}\right) - 8\Li_2\left(\frac{y_1}{y_2}\right)
-8\Li_2\left(\frac{z_1}{s_1}\right) - 8\Li_2\left(\frac{z_1}{s_2}\right)
+8\Li_2\left(\frac{z_1}{z_2} \right)
\right], \\
k^{1/2}I_2^{(1)} &=& - 2s_1^2 - 2s_2^2 + 2s_3^2 - 8s_1s_2 + s_1s_2\Bigl[ \Bigr. \nonumber \\
&& -32\log^2(v_1) - 32\log (v_3)\log (v_1) - 32\log (v_4)\log (v_1) + 4\log (v_1)  \nonumber \\
&&- 8\log^2 (v_3) - 8\log^2 (v_4) - 4\log (v_2) - 4\log (v_3) - 16\log (v_3)(v_4) \nonumber \\
&&\left. -16\Li_2\left(\frac{x_2}{x_1}\right) - 16\Li_2\left(\frac{y_1}{y_2}\right)
-16\Li_2\left(\frac{z_1}{s_1}\right) - 16\Li_2\left(\frac{z_1}{s_2}\right)
+16\Li_2\left(\frac{z_1}{z_2}\right)
\right], \\
k^{1/2}I_3^{(1)} &=& -2s_1^2 + 2s_2^2 + 2s_3^2 -8s_2s_3 +s_2s_3\Bigl[ \Bigr. \nonumber \\
&& -32\log^2(v_1)-32\log (v_3)\log (v_1) -32\log (v_4)\log (v_1) + 4\log (v_1) \nonumber \\
&& -8\log^2(v_3) - 8\log^2(v_4) - 4\log (v_2) - 4\log (v_3) - 16\log (v_3)\log (v_4) \nonumber \\
&& \left.
-16\Li_2\left(\frac{x_2}{x_1}\right) - 16\Li_2\left(\frac{y_1}{y_2}\right)
-16\Li_2\left(\frac{z_1}{s_1}\right) - 16\Li_2\left(\frac{z_1}{s_2}\right)
+16\Li_2\left(\frac{z_1}{z_2}\right)
\right], \\
k^{1/2}I_4^{(1)} &=& 2s_1^2 - 2s_2^2 + 2s_3^2 - 8s_1s_2 + s_1s_3\Bigl[ \Bigr. \nonumber \\
&& -32\log^2 (v_1) - 32\log (v_3)\log (v_1) - 32\log (v_4)\log (v_1) + 4\log (v_1) \nonumber \\
&& -8\log^2 (v_3) - 8\log^2 (v_4) - 4\log (v_2) - 4\log (v_3) - 16\log (v_3)\log (v_4) \nonumber \\
&& \left. 
-16\Li_2\left(\frac{x_2}{x_1}\right) - 16\Li_2\left(\frac{y_1}{y_2}\right)
-16\Li_2\left(\frac{z_1}{s_1}\right) - 16\Li_2\left(\frac{z_1}{s_2}\right)
+16\Li_2\left(\frac{z_1}{z_2}\right)
\right],
\end{eqnarray}
where the following notation was introduced:
\begin{eqnarray}
x_1 &=& \frac{1}{2}(s_1-s_2-Q^2)-\frac{1}{2}\sqrt{k}, \\
x_2 &=& \frac{1}{2}(s_1-s_2-Q^2)+\frac{1}{2}\sqrt{k}, \\
y_1 &=& \frac{1}{2}(s_1+Q^2-s_2)-\frac{1}{2}\sqrt{k}, \\
y_2 &=& \frac{1}{2}(s_1+Q^2-s_2)+\frac{1}{2}\sqrt{k}, \\
z_1 &=& \frac{1}{2}(s_1+s_2+Q^2)-\frac{1}{2}\sqrt{k}, \\
z_2 &=& \frac{1}{2}(s_1+s_2+Q^2)+\frac{1}{2}\sqrt{k}, \\
v_1 &=& \frac{1}{2s_1}(s_1-s_2-Q^2)+\frac{1}{2s_1}\sqrt{k}, \\
v_2 &=& \frac{1}{2s_2}(s_1-s_2+Q^2)+\frac{1}{2s_2}\sqrt{k}, \\
v_3 &=& \frac{1}{2s_1}(s_1+s_2+Q^2)+\frac{1}{2s_1}\sqrt{k}, \\
v_4 &=& \frac{s_1}{Q^2}, \\
v_5 &=& \frac{s_2}{Q^2}, \\
v_6 &=& 1 - \frac{z_1}{z_2}.
\end{eqnarray}
\noindent
We checked, that  all infrared and ultraviolet divergences cancel as should be for vector
interpolating currents.

Now, let us proceed with the physical part of three-point sum rules. The connection
to hadrons in the framework of QCD sum rules is obtained by matching the resulting
QCD expressions of current correlators with spectral representation, derived from
a double dispersion relation at $q^2\leq 0$:
\begin{eqnarray}
\Pi_{\mu\al\be}(p_1^2, p_2^2, q^2) = \frac{1}{(2\pi)^2}
\int\frac{\rho_{\mu\al\be}^{\mathbf{phys}}(s_1, s_2, Q^2)}{(s_1 - p_1^2)(s_2 - p_2^2)}ds_1 ds_2
+ \mbox{subtractions}. \label{disp_phys}
\end{eqnarray} 
Assuming that the dispersion relation (\ref{disp_phys}) is well convergent, the physical
spectral functions are generally saturated by the lowest lying hadronic states plus
a continuum starting at some thresholds $s_1^{th}$ and $s_2^{th}$:
\begin{eqnarray}
\rho_{\mu\al\be}^{\mathbf{phys}}(s_1, s_2, Q^2) &=& \rho_{\mu\al\be}^{\mathbf{res}}(s_1, s_2, Q^2) + \nonumber \\
&& \theta (s_1 - s_1^{th})\cdot\theta (s_2 - s_2^{th})\cdot\rho_{\mu\al\be}^{\mathbf{cont}}(s_1, s_2, Q^2),
\end{eqnarray}
where
\begin{eqnarray}
\rho_{\mu\al\be}^{\mathbf{res}}(s_1, s_2, Q^2) &=& \frac{m_{\rho}^4}{g_{\rho}^2}\Biggl\{ \Biggr.  \nonumber \\
&& \frac{1}{2m_{\rho}^2}P_{\mu}(p_{2\al}p_{2\be}+p_{1\al}p_{1\be})
\left[
F_1 (Q^2) - F_2 (Q^2) + 2\left(1+\frac{Q^2}{2m_{\rho}^2}\right) F_3 (Q^2)
\right] \nonumber \\
&& + p_{2\be}p_{1\al}P_{\mu}\frac{1}{m_{\rho}^2}\left(1+\frac{Q^2}{2m_{\rho}^2} \right)
\left[
F_2 (Q^2) - \left(1+\frac{Q^2}{2m_{\rho}^2}\right)F_3 (Q^2)
\right] \nonumber \\
&& -\frac{1}{m_{\rho}^2}p_{2\al}p_{1\be}P_{\mu} F_3 (Q^2)
+ \frac{1}{2m_{\rho}^2}q_{\mu}(p_{1\al}p_{1\be}-p_{2\al}p_{2\be})
\left[
F_1 (Q^2) + F_2 (Q^2)
\right] \nonumber \\
&& -g_{\al\be}P_{\mu}F_1 (Q^2) + (g_{\al\mu}p_{1\be}+g_{\be\mu}p_{2\al})
\left[
F_1 (Q^2) + F_2 (Q^2)
\right] \nonumber \\
&&  - (g_{\al\mu}p_{2\be}+g_{\be\mu}p_{1\al})\left(1+\frac{Q^2}{2m_{\rho}^2} \right)
\left[
F_1 (Q^2) + F_2 (Q^2) 
\right]
\Biggl. \Biggr\}(2\pi)^2\delta (s_1)\delta (s_2) \nonumber \\ && + \mbox{higher state contributions} 
\end{eqnarray}
The continuum of higher states is modeled by the perturbative 
absorptive part of $\Pi_{\mu\al\be}$, i.e. by $\rho_{\mu\al\be}$. Then, the 
expressions for the form factors $F_i$ can be derived by equating 
the representations for three-point functions $\Pi_{\mu\al\be}$ from (\ref{doubledisp}) and (\ref{disp_phys}).
It is reasonable to consider 3 sum rules: for structures $p_{1\mu}g_{\al\be}+p_{2\mu}g_{\al\be}$,
$p_{1\mu}p_{2\al}p_{1\be}+p_{2\mu}p_{2\al}p_{1\be}$ and $p_{1\be}g_{\al\mu}+p_{2\al}g_{\be\mu}$.
This last step constitutes a formulation of QCD sum rules for our particular problem.

\section{Numerical analysis}

For numerical analysis we used Borel scheme of QCD sum rules. That is, 
to get rid of unknown subtraction terms in (\ref{doubledisp}) we perform
Borel transformation procedure in two variables $s_1$ and $s_2$. The Borel
transform of three-point function $\Pi_{\mu\al\be} (s_1, s_2, q^2)$ is defined as
\begin{eqnarray}
\Phi_{\mu\al\be} (M_1^2, M_2^2, q^2) &\equiv & \hat B_{12}\Pi_{\mu\al\be} (s_1, s_2, q^2) =  \nonumber \\ &&
\lim_{n,m\to\infty}\left\{\left. \frac{s_2^{n+1}}{n!}
\left(-\frac{d}{d s_2}\right)^n \frac{s_1^{m+1}}{m!}
\left(-\frac{d}{d s_1}\right)\right|_{s_1=m M_1^2, s_2 = n M_2^2} \right\}
\Pi_{\mu\al\be} (s_1, s_2, q^2) \nonumber \\  \label{boreltransform}
\end{eqnarray}
Then Borel transformation (\ref{boreltransform}) of (\ref{doubledisp}) and (\ref{disp_phys}) gives 
\begin{eqnarray}
\Phi_{\mu\al\be}^{(\mathbf{pert}|\mathbf{phys})}  (M_1^2, M_2^2, q^2) = 
\frac{1}{(2\pi)^2}\int_0^{\infty}ds_1\int_0^{\infty}ds_2
\exp\left[-\frac{s_1}{M_1^2}-\frac{s_2}{M_2^2}\right]\rho_{\mu\al\be}^{(\mathbf{pert}|\mathbf{phys})} (s_1, s_2, q^2), 
\label{borelcor}
\end{eqnarray}
In what follows we put $M_1^2 = M_2^2 = M^2$. If $M^2$ is chosen to be of order 1 GeV$^2$, 
then the right hand side of (\ref{borelcor}) in the case of physical spectral density will
be dominated by the lowest hadronic state contribution, while the higher state contribution 
will be suppressed.

\begin{figure}[ht]
\vspace*{3cm}
~~~~~~~~~~~~~~~~$F_{1}(Q^2)$
\vspace*{-3cm}
\begin{center}
\includegraphics[scale=1.]{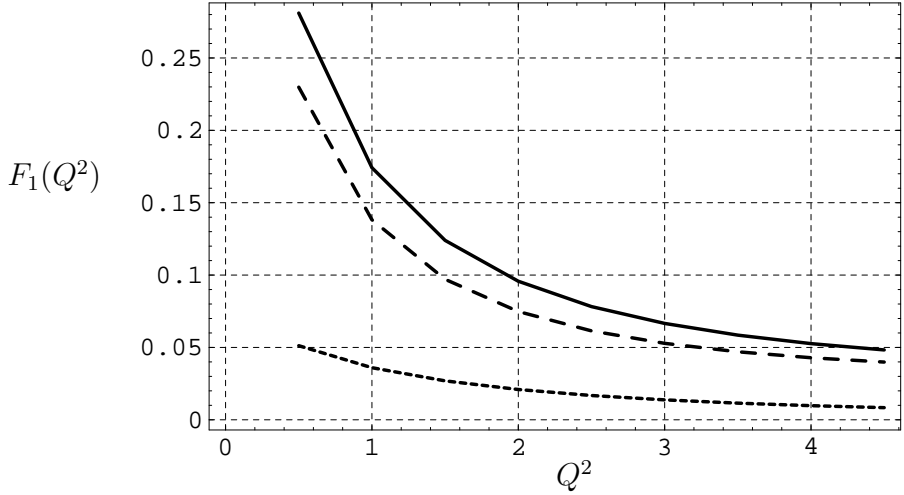} 
\caption{$Q^2$ dependence of $F_1$ electric form factor}
\label{F1plot}
\end{center}
\vspace*{-1.5cm}
~~~~~~~~~~~~~~~~~~~~~~~~~~~~~~~~~~~~~~~~~~~~~~~~~~~~~~~~~~~~~~~~~~~~~$Q^2$
\vspace*{1.cm}
\end{figure}

Now let us recall that our sum rules also receive power corrections proportional to QCD vacuum condensates.
The evaluation of power corrections is simplified if performed directly for the Borel transformed expression
of three-point correlation function and this is the reason why we delayed their discussion up to his moment. 
The quark condensate contribution is known already for a long time and is given by \cite{Ioffe:qb}:
\begin{eqnarray}
\Phi_{\mu\al\be}^{\langle\bar q q\rangle^2}(M^2,M^2,q^2) &=& \frac{4\pi\alpha_s}{81}
\langle 0|\bar q q|0\rangle^2\frac{1}{M^4}\Biggl\{ \Biggr. \nonumber \\
&& P_{\mu}(p_{1\al}p_{1\be}+p_{2\al}p_{2\be})
\left( 
13 - 18\frac{M^2}{Q^2} + 18\frac{M^4}{Q^4} + 2\frac{Q^2}{M^2}
\right) \nonumber \\
&& + P_{\mu}p_{2\be}p_{1\al}
\left(
10 + 36\frac{M^2}{Q^2} - 36\frac{M^4}{Q^4} + 4\frac{Q^2}{M^2}
\right) + 16P_{\mu}p_{1\be}p_{2\al} \nonumber \\
&& + q_{\mu}(p_{1\al}p_{1\be}-p_{2\al}p_{2\be})
\left(
5 - 18\frac{M^2}{Q^2} + 18\frac{M^4}{Q^4} + 2\frac{Q^2}{M^2}
\right) \nonumber \\
&& + (p_{2\be}g_{\al\mu}+p_{1\al}g_{\be\mu})M^2
\left(
64 + 5\frac{Q^2}{M^2} - 2\frac{Q^4}{M^4}
\right) \nonumber \\
&& + 52P_{\mu}g_{\al\be}M^2 + (p_{1\be}g_{\al\mu}+p_{2\al}g_{\be\mu})M^2
\left(
-20 - 8\frac{Q^2}{M^2}
\right) \Biggl. \Biggr\}.
\end{eqnarray}
The correction due to gluon condensate was only partially computed in \cite{Ioffe:qb}, where subset
of diagrams as well as part of Lorentz structures were considered. Taking everything into account 
we get\footnote{See Appendix A for the details of calculation}:
\begin{eqnarray}
\Phi_{\mu\al\be}^{\langle G^2\rangle}(M^2,M^2,q^2) &=&  \frac{\alpha_s}{48\pi}
\langle 0| G_{\rho\sigma}^aG_{\rho\sigma}^a|0\rangle \Biggl\{ \Biggr. \nonumber \\
&& P_{\mu}p_{1\be}p_{2\al}\frac{4}{Q^2} + P_{\mu}p_{1\al}p_{2\be}\frac{2}{M^2}
+ (p_{1\mu}p_{2\be}p_{2\al}+p_{2\mu}p_{1\be}p_{1\al})
\left(
-\frac{4}{Q^2} + \frac{2}{M^2}
\right) \nonumber \\
&& -2g_{\mu\be}p_{2\al} - 2p_{1\be}g_{\mu\al} -\frac{Q^2}{M^2}p_{2\be}g_{\mu\al}
-\frac{Q^2}{M^2}p_{1\al}g_{\mu\be} + 5p_{2\be}g_{\mu\al} +5p_{1\al}g_{\mu\be} \Biggl. \Biggr\}.
\end{eqnarray}  

\begin{figure}[ht]
\vspace*{3cm}
~~~~~~~~~~~~~~~~$F_{2}(Q^2)$
\vspace*{-3cm}
\begin{center}
\includegraphics[scale=1.]{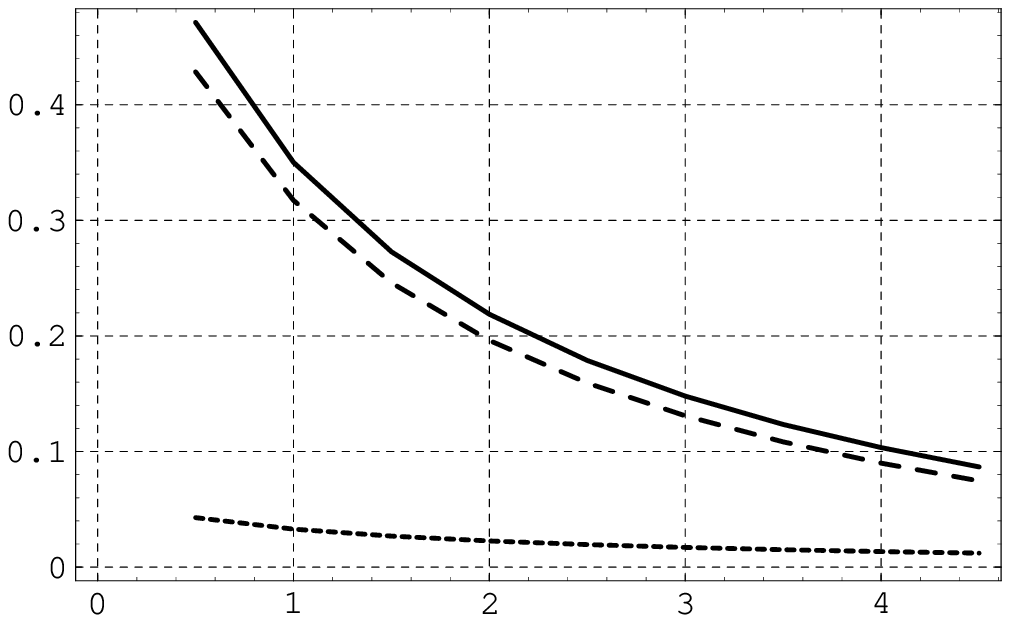} 
\caption{$Q^2$ dependence of $F_2$ magnetic form factor}
\label{F2plot}
\end{center}
\vspace*{-1.5cm}
~~~~~~~~~~~~~~~~~~~~~~~~~~~~~~~~~~~~~~~~~~~~~~~~~~~~~~~~~~~~~~~~~~~~~$Q^2$
\vspace*{1.cm}
\end{figure}

Equating Borel transformed theoretical and physical parts of QCD sum rules we get
\begin{eqnarray}
F_1 (Q^2) &=& \frac{g_{\rho}^2}{8\pi^2}\left(\frac{M^2}{m_{\rho}^2}\right)^2
\exp\left[\frac{2m_{\rho}^2}{M^2}\right]\Biggl\{ \Biggr. \nonumber \\
&& \int_0^{\frac{s_0}{M^2}}dx'\int_0^{x'} dy' s'_3(x'+s'_3)A_2^{(0)}
\exp [-x'] \nonumber \\
&& + \frac{\alpha_s}{4\pi}\int_0^{\frac{s_0}{M^2}}dx'\int_0^{x'} dy' s'_3(x'+s'_3)A_2^{(1)}\exp [-x'] 
 + \frac{640\pi^3}{81M^6}\alpha_s \langle 0|\bar q q|0\rangle^2 \Biggl. \Biggr\},  \\
F_2 (Q^2) &=& \frac{g_{\rho}^2}{8\pi^2}\left(\frac{M^2}{m_{\rho}^2}\right)^2
\exp\left[\frac{2m_{\rho}^2}{M^2}\right]\Biggl\{ \Biggr. \nonumber \\
&& \int_0^{\frac{s_0}{M^2}} dx' \int_0^{x'} dy' 
\left[
s'_1 (y'+s'_3)A_3^{(0)}
- s'_2(y'-s'_3)A_4^{(0)}
- s'_3(x'+s'_3)A_2^{(0)}
\right]
\exp [-x'] \nonumber \\
&& + \frac{\alpha_s}{4\pi}
\int_0^{\frac{s_0}{M^2}} dx' \int_0^{x'} dy' 
\left[
s'_1 (y'+s'_3)A_3^{(1)}
- s'_2(y'-s'_3)A_4^{(1)}
- s'_3(x'+s'_3)A_2^{(1)}
\right]
\exp [-x'] \nonumber \\
&& + \frac{256\pi^3}{81M^6}\alpha_s \langle 0|\bar q q|0\rangle^2 (4-s'_3)
- \frac{\pi^2}{3M^4}\langle 0|\frac{\alpha_s}{\pi}G_{\rho\sigma}^2|0\rangle \Biggl. \Biggr\}, \\
F_3 (Q^2) &=& \frac{g_{\rho}^2}{8\pi^2}\left(\frac{M^2}{m_{\rho}^2}\right)
\exp\left[\frac{2m_{\rho}^2}{M^2}\right]\Biggl\{ \Biggr. \nonumber \\
&& \int_0^{\frac{s_0}{M^2}} dx' \int_0^{x'} dy' 
\left[
2s'_1 A_3^{(0)} + 2s'_2A_4^{(0)} - 2s'_3 A_2^{(0)} - 8s'_1s'_2s'_3 A_1^{(0)}
\right]\exp [-x'] \nonumber \\
&& + \frac{\alpha_s}{4\pi}\int_0^{\frac{s_0}{M^2}} dx' \int_0^{x'} dy' 
\left[
2s'_1 A_3^{(1)} + 2s'_2A_4^{(1)} - 2s'_3 A_2^{(1)} - 8s'_1s'_2s'_3 A_1^{(1)}
\right]\exp [-x'] \nonumber \\
&& - \frac{512\pi^3}{81M^6}\alpha_s 
\langle 0|\bar q q|0\rangle^2
- \frac{2\pi^2}{3M^4 s'_3}\langle 0|\frac{\alpha_s}{\pi}G_{\rho\sigma}^2|0\rangle \Biggl. \Biggr\},
\end{eqnarray}
where the following notation $x' = \frac{x}{M^2}$, $y' = \frac{y}{M^2}$,
$s'_1 = \frac{s_1}{M^2}$, $s'_2 = \frac{s_2}{M^2}$ and $s'_3 = \frac{Q^2}{M^2}$ was introduced. 

\begin{figure}[ht]
\vspace*{3cm}
~~~~~~~~~~~~~~~~$F_{3}(Q^2)$
\vspace*{-3cm}
\begin{center}
\includegraphics[scale=1.]{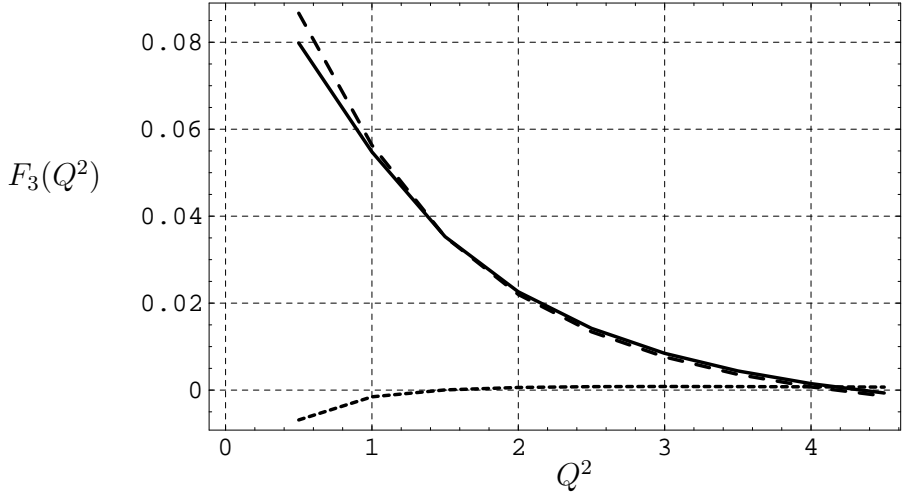} 
\caption{$Q^2$ dependence of $F_3$ quadrupole form factor}
\label{F3plot}
\end{center}
\vspace*{-1.5cm}
~~~~~~~~~~~~~~~~~~~~~~~~~~~~~~~~~~~~~~~~~~~~~~~~~~~~~~~~~~~~~~~~~~~~~$Q^2$
\vspace*{1.cm}
\end{figure}

\begin{figure}[ht]
\vspace*{3cm}
~~~~~~~~~~~~$F_{2}(1~\mbox{GeV}^2)$
\vspace*{-3cm}
\begin{center}
\includegraphics[scale=1.]{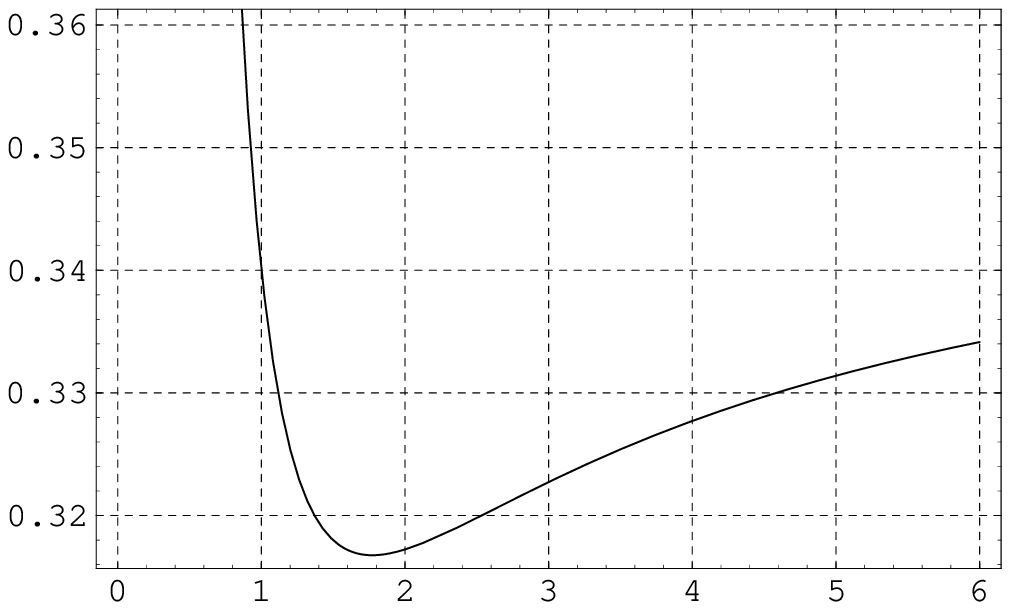} 
\caption{Borel mass $M^2$ dependence of $\rho$-meson magnetic form factor at $Q^2 =1~\mbox{GeV}^2$}
\label{F2plotM2}
\end{center}
\vspace*{-1.5cm}
~~~~~~~~~~~~~~~~~~~~~~~~~~~~~~~~~~~~~~~~~~~~~~~~~~~~~~~~~~~~~~~~~~~~~$M^2$
\vspace*{1.cm}
\end{figure}

Here, for continuum subtraction we used so called "triangle" model. To verify the stability of our results with 
respect to the choice of continuum model we checked, that the usual "square" model gives similar predictions for pion 
electromagnetic form factor provided $s_0\sim 1.5 s_1$ is chosen\footnote{For more information about different continuum
subtraction models see \cite{Ioffe:qb}}. In what follows  we set $s_0 = 2.2$. This value coincides
with one used in the previous analysis \cite{Ioffe:qb} and is in agreement with the value of continuum threshold
employed in the analysis of corresponding two-point QCD sum rules. For the rest of parameters entering our expressions
for form factors we used the following values\footnote{For numerical values of QCD condensates we took central values
of estimates made in \cite{Ioffe:2002ee}}:
\begin{eqnarray}
m_{\rho} &=& 0.77~\mbox{GeV}, \nonumber \\
\langle 0|\frac{\alpha_s}{\pi}G_{\rho\sigma}^2|0\rangle &=& 0.009~\mbox{GeV}^4, \nonumber \\
\alpha_s\langle 0|\bar q q|0\rangle^2 &=& 2.2~ 10^{-4}~\mbox{GeV}^6. \nonumber
\end{eqnarray}
In Fig. \ref{F2plotM2} we plotted the dependence of the $\rho$-meson magnetic form 
factor $F_2 (Q^2)$ from the value of Borel parameter $M^2$ at $Q^2=1~\mbox{GeV}^2$. Similar dependence
was also found for other form factors. To ensure stability of our estimates to the variation of Borel
parameter we fix it value at $M^2 = 2~\mbox{GeV}^2$, belonging to the "stability plateau".

The results for $\rho$-meson electric form factor including NLO corrections are shown in Fig. \ref{F1plot}
(solid line is the sum of LO and NLO contributions, curve with long dashes denotes LO contribution
and curve with short dashes stands for NLO contributions.). 
Similar results for magnetic $F_2 (Q^2)$ and quadrupole $F_3 (Q^2)$ form factors are shown in Figs. \ref{F2plot}
and \ref{F3plot} correspondingly (again solid line is the sum of LO and NLO contributions, curve with long dashes denotes LO contribution
and curve with short dashes stands for NLO contributions.). Unfortunately, our sum rules do not allow us
determine magnetic moment of $\rho$-meson with precision better then already available predictions both from
sum rules \cite{Samsonov:2003hs,Aliev:2003ba} and different quark models \cite{Hawes:1998bz,Hecht:1997uj,deMelo:1997hh}.
At small $Q^2$ our formula simply do not work - at this point we approach physical region of our three-point correlation
function and OPE used in our analysis breaks down. Here one may only conclude, that the value of 
$\rho$-meson magnetic moment is close to $2$ by extrapolating  the ratio of $F_2 (Q^2)/F_1 (Q^2)$ into the region
of small momentum transfer $Q^2$ \cite{Ioffe:qb}. 

Finally, let us following \cite{Ioffe:qb} compare the behaviour of our form factors
with radiative corrections taken into account in the limit $Q^2\to\infty$ 
 with that, predicted by perturbative QCD (pQCD). The asymptotic of $\rho$-meson form factors is
typically discussed in terms of states with given transverse $\ep_T$ and longitudinal $\ep_L$ polarizations
of $\rho$-meson. In Breit system these two descriptions could be related with the help of following
formula:
\begin{eqnarray}
F_{TT} (Q^2) &=& \langle p',\ep_T|j_0^{\mathbf{el}}|p,\ep_T\rangle/2E = F_1 (Q^2), \\
F_{LT} (Q^2) &=& \langle p',\ep_T|j_x^{\mathbf{el}}|p,\ep_L\rangle/2E = \frac{Q}{2m_{\rho}}(F_1 (Q^2)+ F_2 (Q^2)), \\
F_{LL} (Q^2) &=& \langle p',\ep_L|j_0^{\mathbf{el}}|p,\ep_L\rangle/2E = F_1 (Q^2)- \frac{Q^2}{2m_{\rho}^2}F_2 (Q^2)
+ \frac{Q^2}{m_{\rho}^2}\left(1+\frac{Q^2}{4m_{\rho}^2} \right)F_3 (Q^2), 
\end{eqnarray}
where $E$ is the meson energy $E = \sqrt{m_{\rho}^2 +\frac{1}{4}Q^2}$.

Quark counting and chirality conservation lead to the following asymptotic behavior of form factors:
$F_{LL} (Q^2)\sim 1/Q^2$, $F_{LT} (Q^2)\sim 1/Q^3$ and $F_{TT} (Q^2)\sim 1/Q^2$. In terms of electric, magnetic 
and quadrupole form factors we have $F_1 (Q^2)\sim F_2 (Q^2)\sim 1/Q^4$ and $F_3 (Q^2)\sim 1/Q^6$. It is easy to
check that our form factors with radiative corrections included follow this behavior in asymptotic limit, while 
LO results contribute only as power corrections at large momentum transfers.

\section{Conclusion}

We presented the results for $\rho$-meson electromagnetic form factors in the framework of three-point 
NLO QCD sum rules. The radiative corrections are sizeable ($\sim 30\%$ in the case of $F_1$ form factor
and somewhat smaller for two other form factors) and should be taken into account when precision data
on $\rho$-meson form factors become available.

The work  of V.B. was supported in part by Russian Foundation of Basic Research under grant 01-02-16585, 
Russian Education Ministry grant E02-31-96, CRDF grant MO-011-0 and Dynasty foundation.  
The work of A.O. was supported by the National Science Foundation under 
grant PHY-0244853 and by the US Department of Energy under grant DE-FG02-96ER41005.

\appendix

\section{Gluon condensate correction}

Here we present details on the evaluation of power corrections
proportional to gluon condensate. This calculation could be relatively easy
performed directly for the Borel transformed expression of three-point correlation
function. Unfortunately, one of the methods (calculation in coordinate space), we will
discuss below, does not allow us to subtract continuum contribution for gluon condensate
correction. However, the form of the obtained expression leads us to the conclusion,
that this contribution is simply absent in our final result. This conclusion is based
on a fact, that typical continuum contribution may show up only as incomplete 
$\Gamma$-functions. The latter are in fact present in contributions of each separate
diagram, but they are canceling in the sum.     

The gluon condensate contribution to the three-point sum rules is given by diagrams with two
external gluon vacuum fields, depicted
in Fig. \ref{cond}. For calculations we have used the Fock-Schwinger fixed point gauge 
\cite{Fock:1937dy,Schwinger:nm}:
\begin{eqnarray}
x^{\mu}A_{\mu}^a (x) = 0,
\end{eqnarray}
where $A_{\mu}^a$, $a = \{1,2,\ldots ,8\}$ is the gluon field. The use of this gauge allows
us express gauge potential $A_{\mu} (x)$ in terms of field strength and its covariant derivatives at 
origin:
\begin{eqnarray}
A_{\mu}^a (x) = -\frac{1}{2}x_{\nu}G_{\mu\nu}^a (0) - \frac{1}{3}x_{\nu}x_{\al}(D_{\al}G_{\mu\nu})^a (0) + \ldots ...
\label{gluoncoord}
\end{eqnarray} 
or in momentum representation:
\begin{eqnarray}
A_{\mu}^a (k) &=& -\frac{1}{2}i(2\pi)^4G_{\nu\mu} (0)\frac{\partial}{\partial k_{\nu}}\delta^4 (k) \nonumber \\
&& -\frac{1}{3}(2\pi)^4 \left(D_{\al} G_{\nu\mu} (0) \right)^a \frac{\partial^2}{\partial k_{\al}\partial k_{\nu}}
\delta^4 (k) + \ldots \label{gluonmomentum}
\end{eqnarray}

\setlength{\unitlength}{1mm}
\begin{figure}[ph]
\begin{center}
\begin{picture}(150,180)
\put(0,120){\epsfxsize=3cm \epsfbox{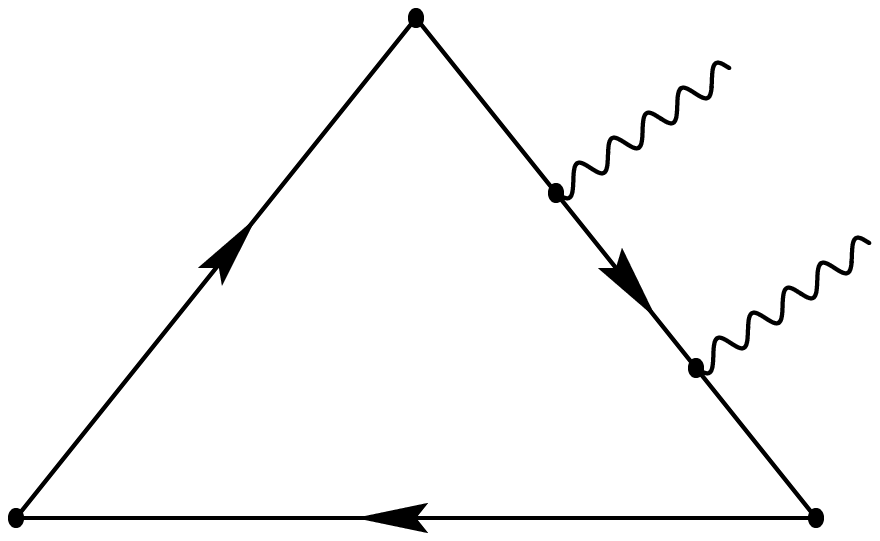}}
\put(75,120){\epsfxsize=3cm \epsfbox{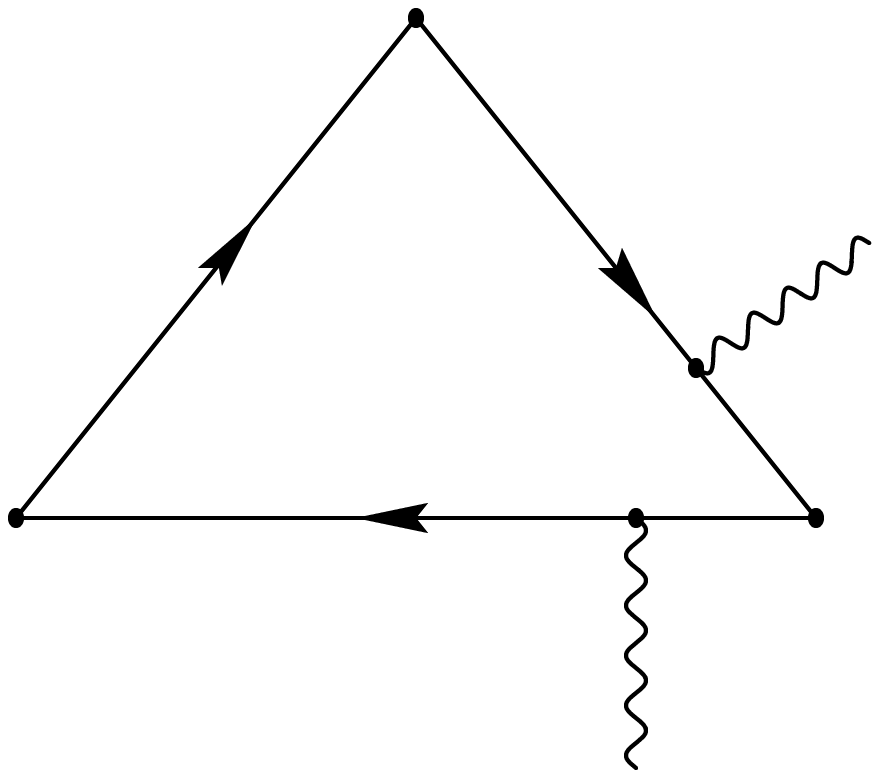}}
\put(0,60){\epsfxsize=3cm \epsfbox{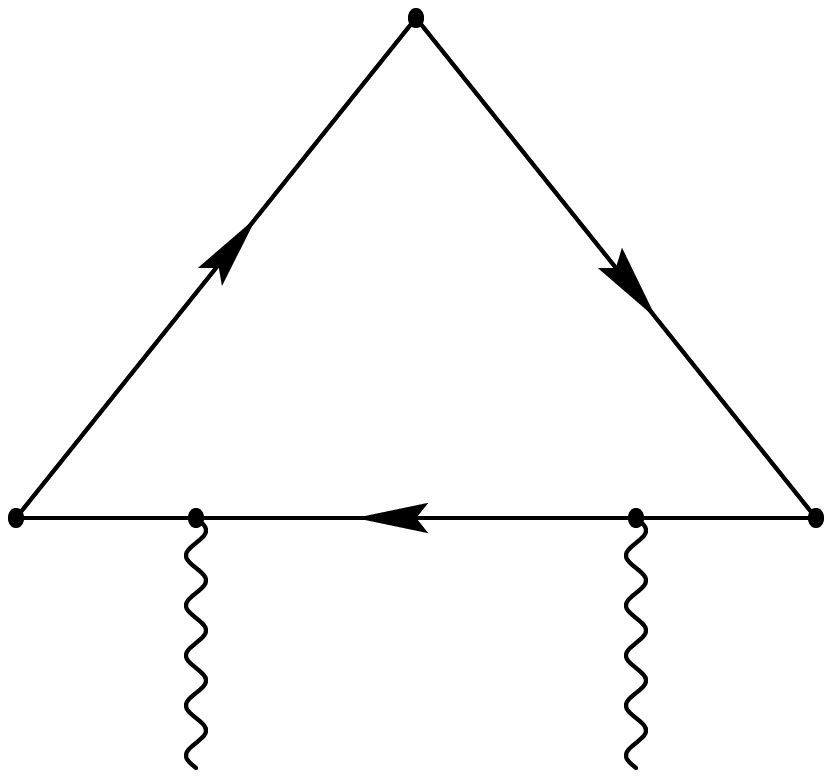}}
\put(75,60){\epsfxsize=3cm \epsfbox{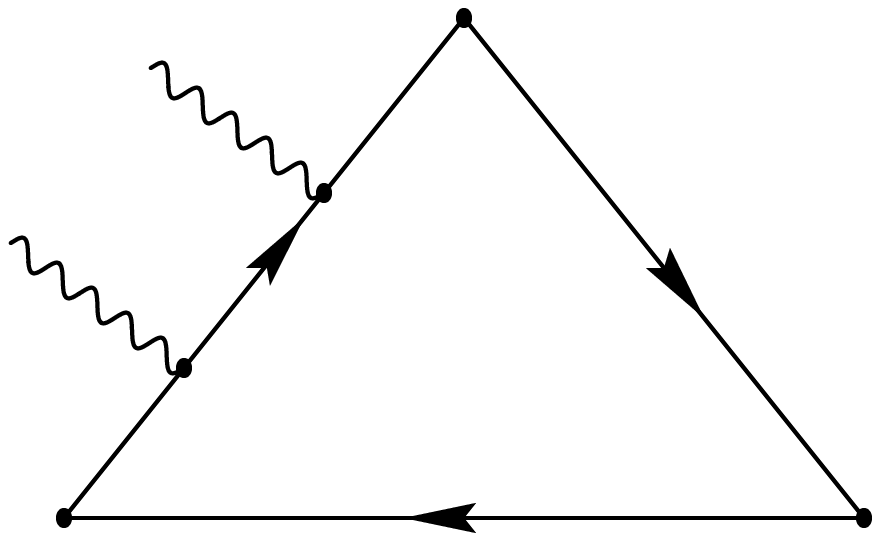}}
\put(0,0){\epsfxsize=3cm \epsfbox{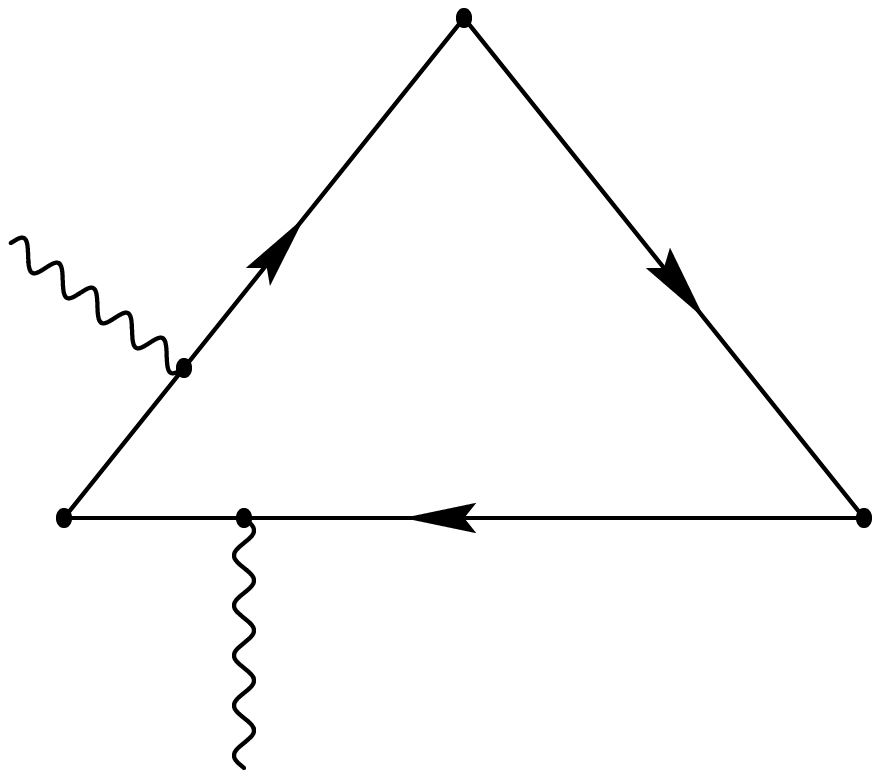}}
\put(75,0){\epsfxsize=3cm \epsfbox{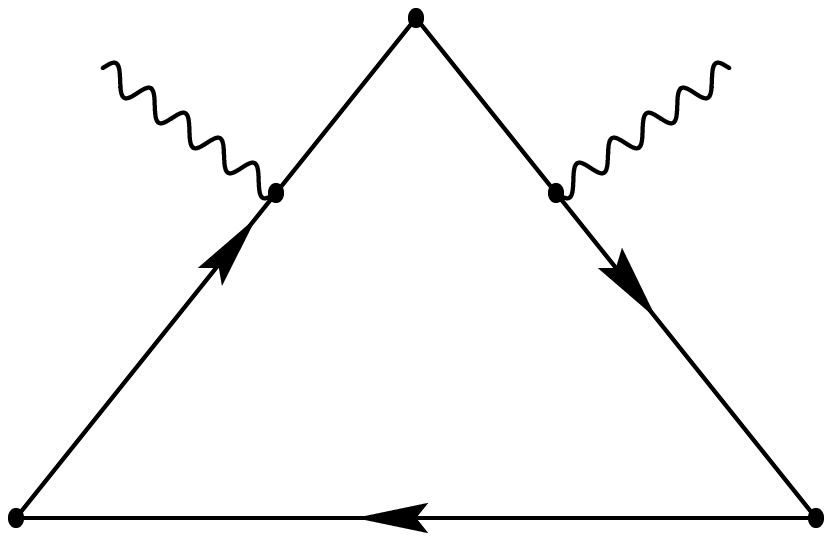}}

\put(0,140){$p_1$}
\put(60,140){$p_2$}
\put(30,135){$k$}
\put(30,125){$\bf a)$}
\put(30,176){$0$}
\put(62,160){$k_1$}
\put(52,172){$k_2$}
\put(6,158){$p_1+k$}
\put(54,148){$p_2+k$}

\put(0,80){$p_1$}
\put(60,80){$p_2$}
\put(30,75){$k$}
\put(30,65){$\bf c)$}
\put(30,116){$0$}
\put(16,62){$k_1$}
\put(43,62){$k_2$}
\put(-3,98){$p_1+k+k_1$}
\put(46,98){$p_2+k-k_2$}

\put(0,20){$p_1$}
\put(60,20){$p_2$}
\put(30,15){$k$}
\put(30,5){$\bf e)$}
\put(30,56){$0$}
\put(16,2){$k_1$}
\put(-2,42){$k_2$}
\put(46,38){$p_2+k$}

\put(75,20){$p_1$}
\put(135,20){$p_2$}
\put(105,15){$k$}
\put(105,5){$\bf f)$}
\put(105,56){$0$}
\put(82,52){$k_1$}
\put(126,52){$k_2$}
\put(78,38){$p_1+k$}
\put(121,38){$p_2+k$}

\put(75,80){$p_1$}
\put(135,80){$p_2$}
\put(105,75){$k$}
\put(105,65){$\bf d)$}
\put(105,116){$0$}
\put(82,112){$k_2$}
\put(73,102){$k_1$}
\put(121,98){$p_2+k$}

\put(75,140){$p_1$}
\put(135,140){$p_2$}
\put(105,135){$k$}
\put(105,125){$\bf b)$}
\put(105,176){$0$}
\put(137,160){$k_2$}
\put(118,122){$k_1$}
\put(81,158){$p_1+k$}

\end{picture}
\end{center}
\caption{The gluon condensate contribution to three-point QCD sum rules. The
directions of $p_1,\; k_1,\; k_2$ momenta are incoming, and that of $p_2$ is
outgoing.}
\label{cond}
\end{figure}
\normalsize

So, basically, the calculation of gluon condensate correction is the ordinary calculation
in background of vacuum gluon fields in the form of (\ref{gluoncoord}) or (\ref{gluonmomentum}). Finally,
vacuum averaging is performed according to rule:
\begin{eqnarray}
\langle 0|G_{\mu\nu}G_{\rho\sigma}^b|0\rangle = \frac{1}{96}\delta^{ab}
\langle 0|G_{\al\be}^cG_{\al\be}^c|0\rangle 
(\delta_{\mu\rho}\delta_{\nu\sigma}-\delta_{\mu\sigma}\delta_{\nu\rho})
\end{eqnarray}
The calculation could be simplified if one uses the expression for quark propagator in background of gluon
vacuum field:
\begin{eqnarray}
S (x,y) &=& \frac{1}{2\pi^2}\frac{\slashchar{r}}{(r^2)^2} 
- \frac{1}{8\pi^2}\frac{r_{\al}}{r^2}\tilde{G}_{\al\be} (0)\gamma_{\be}\gamma_5 \nonumber \\
&& + \left\{
\frac{i}{4\pi^2}\frac{\slashchar{r}}{(r^2)^2}y_{\al}x_{\be}G_{\al\be} (0)
- \frac{1}{192\pi^2}\frac{\slashchar{r}}{(r^2)^2}(x^2y^2 - (x y)^2)
G_{\al\be} (0) G_{\al\be} (0)
\right\} \nonumber \\
&& + \mbox{~~~~operators of higher dimension}, \label{quarkprop}
\end{eqnarray}
where
\begin{eqnarray}
r = x-y,\quad G_{\al\be} = \frac{g}{2}\lambda^aG_{\al\be}^a,\quad 
\tilde{G}_{\al\be} = \frac{1}{2}\ep_{\al\be\mu\nu}G_{\mu\nu} 
\end{eqnarray}
\noindent
The corresponding momentum representation of quark propagator could be easily obtained through Fourier transformation.
Next, for coordinates chosen as indicated in Fig. \ref{cond} from (\ref{quarkprop}) it follows that diagrams 
$a)$ and $d)$ are identically zero. 

Now there are two ways to proceed: first is to perform the whole calculation in momentum representation \cite{Ioffe:qb}
and second way - do the same in coordinate space. Let's consider first calculation in momentum space. It is easy to
find that the contribution of diagrams, where external gluon fields attached to different quark lines, is given by
\begin{eqnarray}
\Pi_{\mu\al\be}^{\mathbf{b+e+f}} (p_1,p_2,q)&=& 
i \frac{g^2 \langle 0|G_{\rho\sigma}^2|0\rangle}{24}\frac{1}{(2\pi )^4}\int d^4 k\frac{1}{k^2 P_1^2 P_2^2}\Biggl\{ \Biggr.
\nonumber \\
&& + \frac{1}{k^2 P_1^2} \mbox{Tr}
[\gamma^{\mu}\widehat k\gamma^{\al}\widehat P_1 \gamma^{\be}\widehat P_2
+ 2\gamma^{\mu}\gamma^{\al}\gamma^{\be}\widehat P_2 P_1\cdot k ] \nonumber \\
&& + \frac{1}{P_1^2 P_2^2} \mbox{Tr}
[\gamma^{\mu}\widehat{P}_2\gamma^{\al}\widehat{k}\gamma^{\be}\widehat{P}_1 
+ 2\gamma^{\mu}\gamma^{\be}\widehat{k}\gamma^{\al}P_1\cdot P_2] \nonumber \\
&& + \frac{1}{k^2 P_2^2}\mbox{Tr}
[\gamma^{\mu}\widehat{P}_1\gamma^{\al}\widehat{P}_2\gamma^{\be}\widehat{k}
+ 2\gamma^{\mu}\widehat{P}_1\gamma^{\al}\gamma^{\be}k\cdot P_2] \Biggl. \Biggr\}, 
\end{eqnarray}
where $P_1 = p_1 + k$ and $P_2 = p_2 + k$.  The integrals entering this expression could be conveniently evaluated
using double spectral representation. For example, 
\begin{eqnarray}
\int d^4k \frac{k_{\mu}P_{1\rho}P_{2\sigma}}{k^2P_1^4P_2^4} &=&
\frac{1}{4}\frac{\partial^2}{\partial p_{1\rho}\partial p_{2\sigma}}
\int d^4 k \frac{k_{\mu}}{k^2P_1^2P_2^2} \nonumber \\
&=& -\frac{i}{4} \frac{\partial^2}{\partial p_{1\rho}\partial p_{2\sigma}}
\int\int d s_1 d s_2 \frac{\pi^2}{\lambda^{3/2}}
\frac{s_2 (s_1-s_2-Q^2)p_{1\mu} + s_1 (s_2-s_1-Q^2)p_{2\mu}}{(s_1-p_1^2)(s_2-p_2^2)} \nonumber \\
\end{eqnarray} 
and similar expressions hold for other two integrals. Finally, for the remained diagram $\bf c)$ the use of 
quark propagator in the background vacuum gluon field (\ref{quarkprop}) allows present this contribution
in the following compact form:
\begin{eqnarray}
\Pi_{\mu\al\be}^{\mathbf c} (p_1,p_2,q) &=& \frac{g^2}{576}\langle 0|G_{\mu\nu}^2|0\rangle
\left[
\frac{\partial^4}{\partial p_{1\rho}\partial p_{2\rho}\partial p_{1\sigma}p_{2\sigma}}
 - \frac{\partial^2}{\partial p_{1\rho}\partial p_{1\rho}}\frac{\partial^2}{\partial p_{2\sigma}\partial p_{2\sigma}}
\right]
\Pi_{\mu\al\be}^{(0)}(p_1,p_2,q),
\end{eqnarray}
where $\Pi_{\mu\al\be}^{(0)}(p_1,p_2,q)$ is LO perturbative contribution to our three-point correlation function.
Performing all differentiations and doing afterwards Borel transform according to
\begin{eqnarray}
\widehat{B}_{P^2}(M^2)\frac{1}{(P^2-m^2)^n} = \frac{1}{(n-1)!}(-1)^n\frac{1}{(M^2)^n}e^{-m^2/M^2}
\end{eqnarray} 
we come to the final expression for gluon condensate contribution presented in the main body of the paper.

Now, let us make a few comments about calculation of gluon condensate contribution within coordinate space 
representation\footnote{For more information see also appendix in \cite{Belyaev:1995ya}}. 
The coordinate space amplitude corresponding to this contribution easily follows from an expression of quark propagator
in the background gluon field (\ref{quarkprop}). However, its Borel transformation is not that trivial. To do it, we
first convert our result into momentum space with the help of the following formula \cite{Belyaev:1992xf}:
\begin{eqnarray}
\frac{1}{\pi^4}\int e^{i p_2 x - i p_1 y}\frac{d^4 xd^4 y}{(x-y)^{2l}y^{2m}x^{2n}}
 = \frac{(-1)^{l+m+n+1}}{4^{l+n+m+1}l!m!n!}\int_0^{\infty}
e^{\tau_1 p_1^2 + \tau_2 p_2^2 + \tau_3 q^2}
\frac{d\tau_1^n d\tau_2^m d\tau_3^l}{(\tau_1\tau_2 + \tau_2\tau_3 + \tau_3\tau_1)^{l+m+n-2}}
\end{eqnarray}
The factors in numerator could be incorporated via:
\begin{eqnarray}
x_{\mu}\to -i\frac{\partial}{\partial p_{2\mu}},\quad y_{\mu}\to i\frac{\partial}{\partial p_{1\mu}}.
\end{eqnarray}
The Borel transformation of the resulting expression is performed with the help of the following formula:
\begin{eqnarray}
\widehat{B}_{P^2}(M^2)e^{-\tau P^2} = \delta (1-\tau M^2)
\end{eqnarray}
The final expression for gluon condensate contribution obtained within this approach coincides with the result
obtained in momentum representation and serves as a check of our result.

\end{document}